# Renormalization of the strongly attractive inverse square potential: Taming the singularity

A. D. Alhaidari[†]

*Saudi Center for Theoretical Physics, P. O. Box 32741, Jeddah 21438, Saudi Arabia*
*Physics Department, King Fahd University of Petroleum & Minerals, Dhahran 31261, Saudi Arabia*

**Abstract**: Quantum anomalies in the inverse square potential are well known and widely investigated. Most prominent is the unbounded increase in oscillations of the particle's state as it approaches the origin when the attractive coupling parameter is greater than the critical value of 1/4. Due to this unphysical divergence in oscillations, we are proposing that the interaction gets screened at short distances making the coupling parameter acquire an effective (renormalized) value that falls within the weak range 0 to 1/4. This prevents the oscillations form growing without limit giving a lower bound to the energy spectrum and forcing the Hamiltonian of the system to be self-adjoint. Technically, this translates into a regularization scheme whereby the inverse square potential is replaced near the origin by another that has the same singularity but with a weak coupling strength. Here, we take the Eckart as the regularizing potential and obtain the corresponding solutions (discrete bound states and continuum scattering states).

## I. INTRODUCTION

The potential function of interest to our study is $V(x) = -\frac{1}{2}\mu x^{-2}$ where $x \geq 0$ and $\mu$ is a dimensionless positive parameter. The strong coupling regime corresponds to values of the coupling constant above the critical value of $\mu = \frac{1}{4}$; below which it is said to be in the weak coupling mode. The strong coupling regime is relevant to the physics of fabrication of nanoscale optical devices [1,2]. It is also the potential used to describe the dipole-bound anions of polar molecules [3] and neutral atoms interacting with a charged wire [4]. In nuclear physics and atomic physics, it is used to model the three-body interaction, including those in Bose-Einstein condensation. Additionally, it has some applications in black hole physics [5]. Moreover, the Efimov effect in three-body systems [6] is attributed to the long-range behavior of an inverse square interaction. Further interest in this potential is found in studies of the renormalization group limit cycle in nonrelativistic quantum mechanics [7-9]. From a theoretical point of view, work on the inverse square potential, which extends over decades, has always been associated with paradoxes. For instance, the solution proposed long ago in [10] failed to give an energy spectrum that is bounded from below. The obtained spectrum is abnormal, as the energy eigenvalue may assume all negative values that extend to minus infinity. Landau and Lifshitz associated the occurrence of this infinite bound state to the classical "fall of the particle to the center" [11,12]. This led to a number of alternative regularization techniques [13-15]. More works on the regularization, renormalization, and self-adjoint extension of this and other singular potentials can be found in [6,8,13,16-24].

---

[†] Present Address: Bahçelievler Mahallesi, Fatih Caddesi, Rauf Orbay Sokak, Çinar Apartmani 1, No. 17, Yalova, Turkey
Phone: +90 (226) 811-4759
Email: haidari@sctp.org.sa

In such singular potential settings, the usual dominance of the kinetic energy term in the Hamiltonian is suppressed by the dominance of the singular interaction term near the singular point. In the case of strongly singular potentials, the wavefunction is ill behaved at the boundary leading to the lack of self-adjointness of the Hamiltonian. One way to resolve this problem is to consider the concept of self-adjoint extension of the Hamiltonian. However, such an extension is not unique in that the energy spectrum and other physical results depend on the chosen extension scheme (e.g., the extension parameter) [22]. A second way to resolve this problem is to use regularization and renormalization approaches. That is, the potential is regularized by introducing a short cutoff distance so that the wavefunction vanishes at this cutoff causing the removal of the undefined behavior at the singular point [19]. However, it has been found that the regularization scheme is very sensitive to the cut-off radius used to regularize the singular behavior of the potential at short distances. Physical quantities, such as the energy spectrum and scattering cross section, change dramatically with small variations in the cut-off radius and hence are not uniquely defined when this cutoff parameter is vanishingly small [8]. The issue of non-equivalence of renormalization and self-adjoint extensions for such singular interaction was also raised recently [23].

Aside from the above approaches there was an alternative method based on particle non-conserving processes [24]. It goes as follows: suppose we consider a particle moving in a singular potential that falls into the singularity and gets absorbed then the singular potential can be interpreted as a particle sink. This proposal is of particular importance because in an experiment with cold atoms moving in an inverse square dipolar potential, it was observed that some of the atoms were absorbed [1].

In intuitive quantum mechanics, an attractive potential ordinarily admits discrete bound states and possibly a continuum of scattering states. However, the inverse square potential violates all our intuition about quantum mechanics. Such discrepancies are referred to as anomalies. An anomaly occurs when a classical symmetry is broken by quantization [25,26]. Manifestation of these anomalies in the inverse square problem takes several forms. The potential has no ground state, and all allowed energies are not quantized. For strong coupling, the Schrödinger equation can be solved, and the boundary conditions satisfied, for every negative energy. The solutions are real and normalizable, and each has an infinite number of zero crossings that become denser as we get closer to the origin. This makes it an intriguing system, and analyzing its paradoxes provides an illuminating introduction to some of the more subtle techniques in contemporary theoretical physics: regularization, renormalization, anomalous symmetry-breaking, self-adjoint extensions, dimensional regularization, ultraviolet cut-off, etc. [22]

Most prominent among the anomalies in the inverse square potential with strong coupling is the unbounded increase in oscillations of the particle's state as it approaches the origin. These infinite number of nodes of the wavefunction indicate that the energy spectrum is not bounded and, thus, the Hamiltonian is not self-adjoint. Now, a measurement of the oscillations in the particle's wavefunction requires a spatial resolution $\Delta x$ of the order of the wavelength. Therefore, near the origin $\Delta x \to 0$ and the quantum mechanical uncertainty relation, $\Delta x \cdot \Delta p \gtrsim \hbar$, entails that $\Delta p \to \infty$, where $p$ is the linear momentum of the particle. This points to a divergences in the particle's momentum/energy, known as "ultraviolet divergence". To eliminate these unphysical divergences we need to renormalize the theory at short distance. Here, this means altering (renormalizing) the value(s) of one or more of the physical parameters of the theory. Now, there is only one parameter in this theory; the strong coupling constant $\mu$. Therefore, the value of $\mu$ (the bare value) needs to be changed near the origin to an effective weak value (the renormalized value) in the range 0 to $\frac{1}{4}$, which we denote by $\tilde{\mu}$. With this new value of the coupling, the rapid oscillations in the wavefunctions near the origin comes to an end. Thus, the energy spectrum will acquire a lower bound and the Hamiltonian of the system regains self-adjointness. To achieve that technically, we are proposing a regularization scheme whereby the inverse square potential is replaced near the origin by another that has the same singularity but with the weak coupling strength $\tilde{\mu}$. We take the Eckart as the regularizing potential and obtain the corresponding discrete bound states and continuum scattering states. The energy spectrum formula, which is valid for highly excited states, is $E_n = E_0 e^{-2n\pi/\eta}$, where $\eta = \sqrt{\mu - 1/4}$ and $E_0$ is the "renormalized ground state" energy that depends on the regularization parameters.

## II. THE REGULARIZATION

The inverse square potential with strong attractive coupling is written as $V(x) = -\frac{1}{2}\mu x^{-2}$, where $x \geq 0$ and $\mu$ is a dimensionless real parameter such that $\mu > \frac{1}{4}$. To deal with the anomalies noted above, we are proposing the following regularization

$$V(x) = \begin{cases} -\dfrac{\mu/2}{x^2} &, \quad x \geq x_0 \\ W(x) &, x_0 > x \geq 0 \end{cases}, \tag{1}$$

where $x_0$ is some small enough (infinitesimal) range. We start by requiring that $W(x)$ has the same inverse square singularity near the origin but with a renormalized (tamed) strength. That is, $\lim_{x \to 0}[W(x)] = -\dfrac{\tilde{\mu}/2}{x^2} + o(x^{-2})$, where $\tilde{\mu}$ is the renormalized short-range coupling parameter such that $\tilde{\mu} < \mu$. Moreover, we impose potential continuity at $x = x_0$; that is, $W(x_0) = -\dfrac{\mu/2}{x_0^2}$. One can show that the following Eckart potential could easily meet these requirements:

$$W(x) = -\frac{\lambda^2}{2}\frac{\gamma + \rho \cosh(\lambda x)}{\sinh^2(\lambda x)}, \tag{2}$$

where $\lambda$, $\gamma$ and $\rho$ are real parameters. The requirement on the $\lim_{x \to 0}[W(x)]$ dictates that $\gamma + \rho = \tilde{\mu}$. Moreover, continuity of the potential at $x_0$ gives $\rho = \mu \cdot \tilde{g}(\lambda x_0)$ where

$$\tilde{g}(y) = \frac{(y^{-1}\sinh y)^2 - \tilde{\mu}/\mu}{(\cosh y) - 1}. \tag{3}$$

In the following section, we obtain the exact bound state solution of the wave equation in the short range region $0 \leq x < x_0$ and in the longer range region $x \geq x_0$. Matching these two solutions and their spatial gradients at $x = x_0$ will produce the energy spectrum and associated eigen-states.

### III. BOUND STATES SOLUTION

In the atomic units $\hbar = m = 1$ and in the short-range region, $0 \le x < x_0$, the wave equation $\left[ -\frac{d^2}{dx^2} + 2W(x) - 2E \right]\psi(x) = 0$ could be rewritten in terms of the variable $z = \cosh(\lambda x)$ as

$$\left[ (1-z^2)\frac{d^2}{dz^2} - z\frac{d}{dz} + \frac{\gamma + \rho z}{1-z^2} - \varepsilon \right]\psi(z) = 0, \tag{4}$$

where $\varepsilon = 2E/\lambda^2$. This homogeneous linear second order differential equation has three regular singularities at $z = \pm 1$ and $z = \infty$ [27]. Thus, by standard results (cf., e.g., Theorem 8.1 on page 156 in [27]), it is possible to transform this differential equation into the hypergeometric equation. Moreover, by an indicial analysis of this differential equation we can write (cf., e.g., the chapter on "Classification of Singularities: Nature of the Solutions in the Neighborhood of a Regular Singularity", page 148 ff in [27])

$$\psi(z) = (z-1)^\alpha (z+1)^\beta f(z), \tag{5}$$

where $\alpha$ and $\beta$ are real with $\alpha > 0$. It is easy to show that

$$\frac{d\psi}{dz} = (z-1)^\alpha (z+1)^\beta \left[ \frac{d}{dz} - \frac{\alpha}{1-z} + \frac{\beta}{1+z} \right] f. \tag{6.1}$$

$$\frac{d^2\psi}{dz^2} = (z-1)^\alpha (z+1)^\beta \left[ \frac{d^2}{dz^2} + 2\left( \frac{\beta}{1+z} - \frac{\alpha}{1-z} \right)\frac{d}{dz} + \frac{\alpha(\alpha-1)}{(1-z)^2} + \frac{\beta(\beta-1)}{(1+z)^2} - \frac{2\alpha\beta}{1-z^2} \right] f \tag{6.2}$$

Consequently, Eq. (4) for $\psi(z)$ becomes an equation for $f(z)$ that reads

$$\left\{ (1-z^2)\frac{d^2}{dz^2} + \left[ (2\beta - 2\alpha) - z(2\beta + 2\alpha + 1) \right]\frac{d}{dz} \right.$$
$$\left. + \frac{1}{1-z}\left[ \alpha(2\alpha - 1) + \tfrac{1}{2}(\gamma + \rho) \right] + \frac{1}{1+z}\left[ \beta(2\beta - 1) + \tfrac{1}{2}(\gamma - \rho) \right] - (\alpha + \beta)^2 - \varepsilon \right\} f(z) = 0 \tag{7}$$

Comparing this with the hypergeometric differential equation

$$\left\{ (1-z^2)\frac{d^2}{dz^2} + \left[ (a+b+1-2c) - z(a+b+1) \right]\frac{d}{dz} - ab \right\} {}_2F_1\left(a,b;c;\tfrac{1-z}{2}\right) = 0, \tag{8}$$

we can identify $f(z)$ with ${}_2F_1\left(a,b;c;\tfrac{1-z}{2}\right)$ provided that:

$$\alpha(2\alpha - 1) = -\tfrac{1}{2}(\gamma + \rho) = -\tfrac{1}{2}\tilde{\mu} \quad \Rightarrow \quad 4\alpha = 1 \pm \sqrt{1 - 4\tilde{\mu}}, \tag{9.1}$$

$$\beta(2\beta - 1) = -\tfrac{1}{2}(\gamma - \rho) = -\tfrac{1}{2}\tilde{\mu} + \rho \quad \Rightarrow \quad 4\beta = 1 \pm \sqrt{1 - 4\tilde{\mu} + 8\rho}, \tag{9.2}$$

$$\varepsilon = -\tfrac{1}{4}(a-b)^2, \tag{9.3}$$

$$c = 2\alpha + \tfrac{1}{2}, \text{ and} \tag{9.4}$$

$$a = \alpha + \beta \pm \sqrt{-\varepsilon}, \quad b = \alpha + \beta \mp \sqrt{-\varepsilon}. \tag{9.5}$$

Since ${}_2F_1(a,b;c;z)$ is invariant under the exchange $a \leftrightarrow b$, then without any loss of generality, we can fix the sign in (9.5) and take $a = \alpha + \beta + \sqrt{-\varepsilon}$ and $b = \alpha + \beta - \sqrt{-\varepsilon}$.

Moreover, one can use the identity: $_2F_1(a,b;c;z) = (1-z)^{c-a-b} {}_2F_1(c-a,c-b;c;z)$ and the $a \leftrightarrow b$ exchange symmetry to show that $\psi(z)$ is unchanged under the map $\beta \to \frac{1}{2} - \beta$. This implies that the solution is independent of the ± sign in the expression for $\beta$ in (9.2) above and, thus, we can fix this sign and choose $4\beta = 1 + \sqrt{1 - 4\tilde{\mu} + 8\rho}$. Consequently, we end up with two independent solutions corresponding to the ± signs in the expression for $\alpha$ in (9.1). Now, reality of $\alpha$ dictates that $\tilde{\mu} \le \frac{1}{4}$. With this upper bound on $\tilde{\mu}$, which makes $0 < \frac{\tilde{\mu}}{\mu} < 1$, and for all values of $x_0$ and $\lambda$ one can show that $\rho > \frac{2}{3}\mu$. Thus, reality of the parameter $\beta$ is guaranteed. Now, if we write $\alpha = \frac{1}{4} \pm \nu$ and $\beta = \frac{1}{4} + \tau$, where $\nu = \sqrt{\frac{1}{4} - \tilde{\mu}}$ and $\tau = \sqrt{\frac{1}{4} - \tilde{\mu} + 2\rho}$, then the wavefunction in the near region $0 \le x < x_0$ becomes a linear combination of the two independent solutions associated with the ± sign in $\alpha$ and could be written as

$$\psi(x) = [1 + \cosh(\lambda x)]^\tau \sqrt{\sinh(\lambda x)} [A_+ \chi_\nu(z) + A_- \chi_{-\nu}(z)], \tag{10}$$

where, $\chi_{\pm\nu}(z) = (z-1)^{\pm\nu} {}_2F_1\left(\pm\nu + \tau + \frac{1}{2} + \sqrt{-\varepsilon}, \pm\nu + \tau + \frac{1}{2} - \sqrt{-\varepsilon}; \pm 2\nu + 1; \frac{1-z}{2}\right)$ and $A_\pm$ are arbitrary normalization constants. One can show that $\chi_{-\nu}(z)$ does not belong to the Hilbert space (i.e., it fails to be square integrable as $x \to 0$) or to the domain of the (self-adjoint) Hamilton operator. Thus, we must set $A_-$ equal to zero.

For the longer range region, $x \ge x_0$, the wave equation is

$$\left[\frac{d^2}{dx^2} + \frac{\mu}{x^2} - k^2\right] \psi(x) = 0, \tag{11}$$

where $k^2 = -2E = -\lambda^2 \varepsilon$. The solution of this equation is well-known (See, for example, Ref. [22]). It is a linear combination of the two modified Bessel functions of the first and second kind and could be written as

$$\psi(x) = \sqrt{kx} \left[B_+ I_{i\eta}(kx) + B_- K_{i\eta}(kx)\right], \tag{12}$$

where $\eta = \sqrt{\mu - \frac{1}{4}}$ and $B_\pm$ are normalization constants. However, $\sqrt{kx} I_{i\eta}(kx)$ diverges at infinity and thus $B_+$ must vanish. Therefore, we can finally write the complete bound state wavefunction as

$$\psi(x) = A \begin{cases} [1 + \cosh(\lambda x)]^\tau \sqrt{\sinh(\lambda x)} \chi_\nu(z) &, 0 \le x < x_0 \\ B\sqrt{kx} K_{i\eta}(kx) &, x \ge x_0 \end{cases} \tag{13}$$

where $A$ and $B$ are arbitrary normalization constants. Now, we impose continuity of the wavefunction and its spatial derivative at $x = x_0$ near the origin. These two conditions should give us the constant $B$ and the energy spectrum $\{E_n\}$. The overall wavefunction normalization gives $A$. Specifically, continuity of the wavefunction gives

$$B = \sqrt{\sinh(\lambda x_0)/kx_0} \, (z_0 + 1)^\tau \chi_\nu(z_0)/K_{i\eta}(kx_0), \tag{14}$$

where $z_0 = \cosh(\lambda x_0) \approx 1$. Expanding near the origin where $\lambda x_0 \ll 1$ and $kx_0 \ll 1$, we obtain

$$B \approx 2^{\tau-\nu} \eta |\Gamma(i\eta)| \sinh(\eta\pi) \sqrt{\tfrac{\lambda}{k}} \, (\lambda x_0)^{2\nu} \Big/ \pi \cos\left[\eta \ln\left(\tfrac{1}{2} kx_0\right) - \arg \Gamma(i\eta)\right], \tag{15}$$

where we have used the expansion

$$K_\alpha(x) = \frac{-\pi/2\alpha}{\sin(\alpha\pi)} \left\{ \frac{(x/2)^\alpha}{\Gamma(\alpha)} \left[1 + \frac{(x/2)^2}{1(1+\alpha)} \left[1 + \frac{(x/2)^2}{2(2+\alpha)} \left[1 + \frac{(x/2)^2}{3(3+\alpha)} \left[1 + \ldots \right]\right]\right]\right] \right.$$
$$\left. + \frac{(x/2)^{-\alpha}}{\Gamma(-\alpha)} \left[1 + \frac{(x/2)^2}{1(1-\alpha)} \left[1 + \frac{(x/2)^2}{2(2-\alpha)} \left[1 + \frac{(x/2)^2}{3(3-\alpha)} \left[1 + \ldots \right]\right]\right]\right] \right\} \quad (16)$$

and the identity $x^{i\alpha}/\Gamma(i\alpha) = \frac{1}{|\Gamma(i\alpha)|} \exp i[\alpha \ln(x) - \arg \Gamma(i\alpha)]$. To calculate the gradient of the wavefunction we use the following differential formulas

$$\frac{d}{dz} {}_2F_1(a,b;c;z) = \frac{ab}{c} {}_2F_1(a+1,b+1;c+1;z) \text{, and} \quad (17.1)$$

$$\frac{d}{dx} K_\alpha(x) = -\frac{1}{2}[K_{\alpha+1}(x) + K_{\alpha-1}(x)]. \quad (17.2)$$

Continuity of the gradient of the wavefunction at $x = x_0$ and the use of the expression for the constant $B$ obtained in (15) above dictate that

$$\tan\left[\eta \ln\left(\tfrac{1}{2}kx_0\right) - \arg \Gamma(i\eta)\right] = -2\nu/\eta. \quad (18)$$

Since the tangent determines the angle modulo an integer multiple of $\pi$, then we can write $\eta \ln\left(\tfrac{1}{2}kx_0\right) - \arg \Gamma(i\eta) = -\tan^{-1}(2\nu/\eta) + n\pi$, where $n = 0, \pm 1, \pm 2, \ldots$. Therefore, the wave number $k$ is discretized as

$$k_n = \frac{2}{x_0} \exp\left\{\frac{1}{\eta}\left[n\pi + \arg \Gamma(i\eta) - \tan^{-1}(2\nu/\eta)\right]\right\}. \quad (19)$$

We should recall that in the derivation of (19) we assumed that $kx_0 \ll 1$. Since the value of $\left[\arg \Gamma(i\eta) - \tan^{-1}(2\nu/\eta)\right]$ falls within the range $\left[-\pi, \tfrac{3}{2}\pi\right]$, then for all values of $\mu$ larger than, but not too close to, $\tfrac{1}{4}$ this requires that the integer $n$ be chosen negative and large. Therefore, the energy spectrum formula associated with the regularized potential (1) is finally written as

$$E_n = -\frac{1}{2}k_n^2 = -\frac{2}{x_0^2}\exp\left\{-\frac{2}{\eta}\left[n\pi - \arg \Gamma(i\eta) + \tan^{-1}(2\nu/\eta)\right]\right\} = E_0 e^{-2n\pi/\eta}, \quad (20)$$

where $n$ is large enough and positive corresponding to the wavefunction (13) with $n$ nodes and

$$E_0(\tilde{\mu}, x_0) = -\frac{2}{x_0^2}\exp\left\{\frac{2}{\eta}\left[\arg \Gamma(i\eta) - \tan^{-1}(2\nu/\eta)\right]\right\}, \quad (21)$$

is the "renormalized ground state" energy. It is not the true ground state since the energy spectrum formula (20) holds only asymptotically, viz., for large enough $n$, and therefore furnishes only the part of the discrete spectrum associated with the highly excited states. This energy spectrum depends on the infinitesimal regularization range $x_0$ and on the renormalized short-range coupling parameter $\tilde{\mu}$. It is interesting to compare the distribution of the energy levels in this spectrum, which is exponentially decaying, with other known spectra such as the harmonic oscillator, $E_n \sim n$; the Coulomb, $E_n \sim 1/n^2$; and the Morse, $E_n \sim n^2$. The spectrum formula (20) indicates that as $\mu$ increases the renormalized ground state energy becomes lower, which expands the energy spectrum band, until $\mu$ reaches a value where the $\arg\Gamma(i\eta)$ experiences a jump of $2\pi$ at which $E_0$ changes abruptly by a factor of $e^{-4\pi/\eta}$ to a higher level creating in effect an abrupt shrinking in the energy spectrum band. Figure 1 illustrates this behavior of $E_0$ with increasing $\mu$. In the following section, we obtain the continuum scattering wavefunction and associated phase shift.

## IV. SCATTERING SOLUTION

In the short-range region, $0 \leq x < x_0$, the solution of the wave equation for positive scattering energy that vanishes at the origin is identical to the wavefunction (10) but with $\sqrt{-\varepsilon} \to i\sqrt{\varepsilon}$, that is

$$\psi(x) = A\,(z^2-1)^{1/4}(z+1)^\tau (z-1)^\nu \times$$
$$\phantom{\psi(x)=}{}_2F_1\!\left(\nu+\tau+\tfrac{1}{2}+i\sqrt{\varepsilon},\nu+\tau+\tfrac{1}{2}-i\sqrt{\varepsilon};2\nu+1;\tfrac{1-z}{2}\right) \qquad (22)$$

On the other hand, in the longer range region, $x \geq x_0$, the solution of wave equation is a linear combination of the two Hankel functions and could be written as [22]

$$\psi(x) = \sqrt{kx}\left[B_+ H^+_{i\eta}(kx) + B_- H^-_{i\eta}(kx)\right], \qquad (23)$$

where $k^2 = 2E = \lambda^2 \varepsilon$ and $H^\pm_\alpha(x) = J_\alpha(x) \pm iY_\alpha(x)$. Therefore, the complete scattering wavefunction becomes

$$\psi(x) = \begin{cases} A\sqrt{\sinh(\lambda x)}\left[\cosh(\lambda x)+1\right]^\tau\left[\cosh(\lambda x)-1\right]^\nu \times \\ {}_2F_1\!\left(\nu+\tau+\tfrac{1}{2}+i\sqrt{\varepsilon},\nu+\tau+\tfrac{1}{2}-i\sqrt{\varepsilon};2\nu+1;\tfrac{1-\cosh(\lambda x)}{2}\right), & 0 \leq x < x_0 \\ \sqrt{kx}\left[B_+ H^+_{i\eta}(kx) + B_- H^-_{i\eta}(kx)\right], & x \geq x_0 \end{cases} \qquad (24)$$

For a normalized flux incident from right, the scattering boundary condition at infinity gives

$$\lim_{x\to\infty}\psi(x) = e^{-i(kx+\delta)} + R\,e^{+i(kx+\delta)} = e^{-i\delta}\left[e^{-ikx} + R\,e^{2i\delta}e^{+ikx}\right], \qquad (25)$$

where $R$ is the reflection amplitude, $e^{2i\delta}$ is the scattering matrix, and $\delta$ is the scattering phase shift that depends on the energy and parameters of the theory. Conservation of probability dictates that $R=1$. Using the asymptotic formula [28]

$$\lim_{x\to\infty}\sqrt{kx}\,H^\pm_{i\eta}(kx) = \sqrt{\tfrac{2}{\pi}}\,e^{\pm\pi\eta/2}e^{\pm i(kx-\pi/4)}, \qquad (26)$$

we find that $B_\pm = \sqrt{\tfrac{\pi}{2}}\,e^{\mp\pi\eta/2}e^{\pm i(\delta+\pi/4)}$. Continuity of the wavefunction and its spatial derivative at $x=x_0$ will determine the remaining two unknowns $A$ and $\delta$ giving (for $\lambda x_0 \ll 1$ and $kx_0 \ll 1$):

$$A = \frac{2^{\nu-\tau}\sqrt{2\pi k/\lambda}\,(\lambda x_0)^{-2\nu}}{\eta\,|\Gamma(i\eta)|\sinh(\eta\pi)}\left\{e^{\pi\eta/2}\sin\!\left[\eta\ln\!\left(\tfrac{1}{2}kx_0\right) - \arg\Gamma(i\eta) + \delta + \tfrac{\pi}{4}\right]\right.$$
$$\left. - e^{-\pi\eta/2}\sin\!\left[\eta\ln\!\left(\tfrac{1}{2}kx_0\right) - \arg\Gamma(i\eta) - \delta - \tfrac{\pi}{4}\right]\right\} \qquad (27)$$

$$\tan\!\left(\delta + \tfrac{\pi}{4}\right) = \frac{\eta}{2\nu}\tanh(\eta\pi/2)\,\frac{1 - \tfrac{2\nu}{\eta}\tan\!\left[\eta\ln\!\left(\tfrac{1}{2}kx_0\right) - \arg\Gamma(i\eta)\right]}{1 + \tfrac{\eta}{2\nu}\tan\!\left[\eta\ln\!\left(\tfrac{1}{2}kx_0\right) - \arg\Gamma(i\eta)\right]}. \qquad (28)$$

In the derivation of these results, we used the following approximations

–7–

$$\lim_{x \to 0}\left[ H_{i\eta}^{\pm}(x) \right] = \frac{\mp i}{\eta \sinh(\eta\pi)} \left[ e^{\pm\eta\pi} \frac{(x/2)^{i\eta}}{\Gamma(i\eta)} + \frac{(x/2)^{-i\eta}}{\Gamma(-i\eta)} \right], \tag{29.1}$$

$$\lim_{x \to 0}\left[ \frac{d}{dx} H_{i\eta}^{\pm}(x) \right] = \frac{\pm 1}{x \sinh(\eta\pi)} \left[ e^{\pm\eta\pi} \frac{(x/2)^{i\eta}}{\Gamma(i\eta)} - \frac{(x/2)^{-i\eta}}{\Gamma(-i\eta)} \right]. \tag{29.2}$$

Expression (28) is a very appealing result when compared to similar findings in the literature [29]. Moreover, it shows clearly that the scattering matrix diverges at exactly the bound states energies, which are obtained by Eq. (18).

## V. CONCLUSION AND DISCUSSION

Classically, particles subject to a force derived from the inverse square potential fall to the origin with an infinite velocity. In quantum theory, however, the wavefunction oscillates indefinitely while spiraling down to the origin, leading to indeterminacy of the boundary condition at the singularity. The strong divergence of the potential at the singular point governs the leading physical behavior of the system. Consequently, the standard methods of regular quantum mechanics are supplemented by regularization or renormalization schemes − a powerful tool for studying the behavior of physical theories at various length scales. In this article, we were able to tame this behavior near the origin for the inverse square potential by renormalizing the coupling parameter into the weak regime. As a result, we recover the usual intuitive quantum mechanical description of the system that retained most of the features of the bare (un-renormalized) system such as the rapid oscillation of the wavefunction near the origin albeit with bounds. The exact bound states solution of the problem is obtained in equations (13), (15) and (20) above. Moreover, in section IV we obtained the scattering matrix and associated wavefunction.

The potential regularization scheme in (1) could be generalized as follows ($j = 0, 1, .., N-1$ and $N$ is some positive integer):

$$V(x) = \begin{cases} -\dfrac{\mu/2}{x^2} & , \quad x \geq x_0 \\ W_j(x) & , x_j > x \geq x_{j+1} \end{cases}, \tag{30}$$

where $x_N = 0$, $\lim_{x \to 0}\left[ W_j(x) \right] = -\dfrac{\tilde{\mu}_j/2}{x^2}$, $W_j(x_j) = W_{j-1}(x_j)$, with $\tilde{\mu}_j \in \left[0, \tfrac{1}{4}\right]$ and $W_{-1}(x) \equiv -\dfrac{\mu/2}{x^2}$.

**ACKNOWLEDGMENTS:** The generous support provided by the Saudi Center for Theoretical Physics (SCTP) is highly appreciated. We also acknowledge partial support by King Fahd University of Petroleum and Minerals under group projects number RG1109-1 & RG1109-2. We are grateful to the anonymous Referee for pointing out some typos and errors in the original version of the paper and for suggesting changes that resulted in improving the presentation.

**FIGURE CAPTION**

**Fig. 1**: Variation of the renormalized ground state energy (in units of $x_0^{-2}$) as the coupling constant $\mu$ increases for a given renormalized coupling parameter $\tilde{\mu}$. The horizontal dashed line is at $E_0 = -2/x_0^2$. See text for clarification of the jumps in the graph.

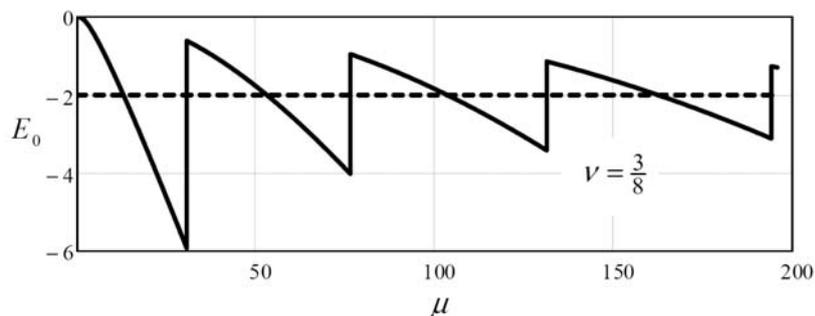

**Fig. 1**